\documentstyle[preprint,pra,aps]{revtex}
\begin{document}
\draft
\title{\bf The nuclear Schiff moment and time invariance violation in atoms}
\author{V.V. Flambaum and J.S.M. Ginges}
\address{School of Physics, University of New South Wales,
Sydney 2052, Australia}
\date{\today}
\maketitle

\tightenlines
\begin{abstract}

Parity and time invariance violating ($P,T$-odd) nuclear forces
produce $P,T$-odd nuclear moments. In turn, these moments
can induce electric dipole moments (EDMs) in atoms through the
mixing of electron wavefunctions of opposite parity.
The nuclear EDM is screened by atomic electrons.
The EDM of an atom with closed electron subshells is induced by the
nuclear Schiff moment.
Previously the interaction with the Schiff moment has been defined
for a point-like nucleus.
No problems arise with the calculation of the electron matrix element
of this interaction as long as the electrons are considered to be
non-relativistic.
However, a more realistic model obviously involves
a nucleus of finite-size and relativistic electrons.
In this work we have calculated the finite nuclear-size and relativistic
corrections to the Schiff moment.
The relativistic corrections originate from the electron
wavefunctions and are incorporated into a ``nuclear'' moment,
which we term the local dipole moment.
For ${^{199}{\rm Hg}}$ these corrections amount to $\sim 25\%$.
We have found that the natural generalization of the
electrostatic potential of the Schiff moment
for a finite-size nucleus corresponds to an electric field
distribution which, inside the nucleus, is well approximated as
constant and directed along the nuclear spin, and outside the
nucleus is zero. Also in this work the ${^{239}{\rm Pu}}$ atomic EDM
is estimated.

\end{abstract}
\vspace{1cm}

\section{Introduction}

The best limit on parity and time invariance violating ($P,T$-odd)
nucleon-nucleon interactions (as well as quark-quark $P,T$-odd
interactions) has been obtained from the
measurement of the ${^{199}{\rm Hg}}$ electric dipole moment (EDM)
\cite{fortson}.
The mechanism of this EDM generation is the following.
$P,T$-odd nuclear forces create $P,T$-odd nuclear moments, e.g. the
EDM and Schiff moment (SM). According to the Schiff theorem
\cite{ramsey,schiff,sandars}, the EDM of a
point-like nucleus is completely screened by atomic electrons,
so it cannot be measured. However, the electrostatic interaction
between atomic electrons and the nuclear Schiff moment induces
an atomic EDM.

The electrostatic potential produced by the Schiff moment
is usually presented in the form \cite{FKS84}
\begin{equation}
\label{olddef}
\varphi ({\bf R})=4\pi {\bf S}\cdot \mbox{\boldmath$\nabla$}\delta ({\bf R}),
\end{equation}
where ${\bf S}$ is the Schiff moment (vector), $\delta ({\bf R})$
is a delta-function.
The contact interaction $-e\varphi$ mixes $s$- and $p$-wave electron
orbitals and produces EDMs in atoms;
for example the atomic EDM induced in an atom with a single electron
in state $ns$ has the form
\begin{equation}
\label{atomedmform}
{\bf d}_{\rm atom}=2\sum_{m}\frac{\langle ns|-e\varphi ({\bf R})|mp\rangle
\langle mp|-e{\bf R}|ns\rangle}{E_{ns}-E_{mp}} \ .
\end{equation}
The expression (\ref{olddef}) is consistently defined for
non-relativistic electrons.
Using integration by parts, it is seen that the matrix element
$\langle s|-e\varphi|p\rangle$ is finite,
\begin{equation}
\label{eq:nonrelS}
\langle s|-e\varphi|p\rangle =4\pi e{\bf S}\cdot
(\mbox{\boldmath$\nabla$}
\psi _{s}^{\dagger}
\psi _{p})_{R=0}={\rm constant}\ .
\end{equation}

However, atomic electrons near the nucleus are ultra-relativistic,
the ratio of the kinetic or potential energy to $mc^{2}$ in heavy atoms
is about $100$.
For the solution of the Dirac equation,
$(\nabla \psi _{s}^{\dagger}\psi _{p})_{R\rightarrow 0}\rightarrow \infty$
for a point-like nucleus. Usually this problem is solved by a cut-off of
the electron wavefunctions at the nuclear surface. However,
even inside the nucleus
$\nabla \psi_{s}^{\dagger}\psi_{p}$ varies significantly,
$\approx Z^{2}\alpha^{2}$, where $\alpha$ is the fine-structure constant,
$Z$ is the nuclear charge. In Hg ($Z=80$), $Z^{2}\alpha^{2}=0.34$.
Recently, proposals have been made to measure EDMs of very heavy atoms
like Ra ($Z=88$) \cite{AFS,SAF,flamra1,flamra2} and
Pu ($Z=94$) \cite{hayes} where $P,T$-violating
nuclear moments and the resulting atomic EDMs are very strongly enhanced.

A consistent treatment of the Schiff moment is needed especially
because the Schiff moment itself is defined as the difference of two
approximately equal terms (see Eq. (\ref{s})).
The aim of the present
work is to develop a consistent theory of the nuclear Schiff moment, and
the atomic EDM it induces, which properly takes into account the relativistic
character of the electron wavefunctions inside the nucleus.

\begin{figure}[b]
\centerline{\psfig{file=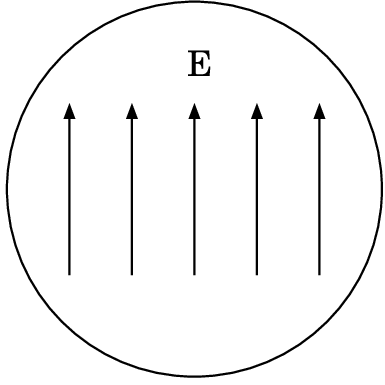,clip=}}
\caption{Constant electric field ${\bf E}$ inside the nucleus produced by
the Schiff moment.
${\bf E}$ is directed along the nuclear spin ${\bf I}$.}
\label{fig:schiff}
\end{figure}

For relativistic electrons we should introduce a finite-size Schiff
moment potential. In this paper we show that the natural generalization
of the Schiff moment potential for a finite-size nucleus is
\begin{equation}
\label{phigen}
\varphi ({\bf R})=-\frac{3 {\bf S}\cdot {\bf R}}{B}n(R)\ ,
\end{equation}
where $B=\int n(R)R^{4}dR\approx R_{N}^{5}/5$,
$R_{N}$ is the nuclear radius,
and $n(R)$ is a smooth function which is $1$ for $R<R_{N}-\delta$ and
$0$ for $R>R_{N}+\delta$;
$n(R)$ can be taken as proportional to the nuclear density.
This potential (\ref{phigen}) corresponds to a constant electric field
inside the nucleus (see Fig. \ref{fig:schiff}) which can be
produced by $P,T$-odd nuclear forces or by an intrinsic EDM of
an external nucleon.
This expression has no singularities and may be used in relativistic
atomic calculations.

A more accurate treatment requires the calculation of a new nuclear
characteristic which we call the
{\it local dipole moment} (LDM).
This moment takes into account relativistic corrections to the nuclear
Schiff moment which originate from the electron wavefunctions.
So in the non-relativistic limit, $Z\alpha \rightarrow 0$, the LDM $L=S$.
For ${^{199}{\rm Hg}}$,
$L\approx S(1-0.8 Z^{2}{\alpha}^{2})\approx 0.75S$.
When considering the interaction of atomic electrons with the LDM we
define it as placed at the center of the nucleus, that is the
electrostatic potential is
\begin{equation}
\varphi ({\bf R})=4\pi {\bf L}\cdot \mbox{\boldmath$\nabla$}
\delta ({\bf R}) \ .
\end{equation}

This paper is organized in the following manner.
In Section \ref{s:schiffpot} we derive a general expression for the
dipole component of the $P,T$-odd electrostatic potential
inside the nucleus.
In Section \ref{s:me} we take the electronic matrix element of this
potential and show that it is related to the nuclear Schiff moment.
In this section the electronic and nuclear problems are separated.
It is shown that it is convenient to define a ``nuclear'' moment
(the local dipole moment) which is the
nuclear Schiff moment with higher-order (relativistic) corrections which
originate from the electron wavefunctions.
In Section \ref{s:ldm} we calculate various nuclear LDMs which arise due to
$P,T$-odd nuclear forces; we calculate the
contribution of an external proton and that of core protons to the
LDM of a spherical nucleus,
and we calculate the collective LDM of an octupole-deformed nucleus.
Then in Section \ref{s:efield} we calculate the electric field distribution
associated with the nuclear Schiff moment.
In Section \ref{s:atomicedm} we estimate the size of the atomic EDM
induced in ${^{239}{\rm Pu}}$.


\section{The Schiff contribution to the nuclear electrostatic potential}
\label{s:schiffpot}

The nuclear electrostatic potential with electron screening taken into
account can be presented in the following form
(see, e.g., Appendix of Ref. \cite{SAF} for derivation):
\begin{equation}
\label{phi}
\varphi ({\bf R})=\int \frac{e\rho ({\bf r})}{|{\bf R}-{\bf r}|}d^{3}r
+\frac{1}{Z}({\bf d}\cdot \mbox{\boldmath$\nabla$})
\int \frac{\rho({\bf r})}{|{\bf R}-{\bf r}|}d^{3}r \ ,
\end{equation}
where $e{\rho}$ is the nuclear charge density, $\int \rho d^{3}r=Z$,
and ${\bf d}=e\int \rho {\bf r}d^{3}r\equiv e\langle {\bf r}\rangle$ is the
nuclear EDM. The second term cancels the dipole long-range electric
field in the multipole expansion of $\varphi ({\bf R})$.
The Coulomb potential $\frac{1}{|{\bf R}-{\bf r}|}$ can be expanded in
terms of Legendre polynomials
\begin{equation}
\label{leg}
\frac{1}{|{\bf R}-{\bf r}|}=\sum _{l}
\frac{r_{<}^{l}}{r_{>}^{l+1}}P_{l}(\cos\theta),
\end{equation}
where $r_{<}\ \big( r_{>}\big)$ is
${\rm min}[r,R]\ \big({\rm max}[r,R]\big)$.
The $P,T$-odd part of the potential (\ref{leg}) originates from the odd
harmonics $l$. The third harmonic $l=3$ corresponds to the octupole
field which has been considered in \cite{murray}.
The contribution of the $l=3$ term is usually small (in ${^{199}{\rm Hg}}$,
which has nuclear spin $I=1/2$, it vanishes).
Higher $l$ always gives negligible contributions. Therefore, we
concentrate on $l=1$ (it may be presented as
$\frac{{\bf r}\cdot{\bf R}}{R^{3}}\Theta (R-r)+
\frac{{\bf r}\cdot{\bf R}}{r^{3}}\Theta (r-R)$,
where $\Theta(r-R)=1$ for $r>R$ and $\Theta(r-R)=0$ for $r<R$):
\begin{equation}
\label{varphi1}
\varphi ^{(1)}({\bf R})=\frac{e{\bf R}}{R^{3}}\cdot
\Big[
\int_{0}^{R}
{\bf r}\rho({\bf r})d^{3}r
\Big] +
e{\bf R}\cdot \Big[
\int_{R}^{\infty}
\frac{\bf r}{r^{3}}\rho ({\bf r})d^{3}r \Big] -
\frac{e\langle{\bf r}\rangle \cdot {\bf R}}{ZR^{3}}
\int_{0}^{R}\rho ({\bf r})d^{3}r \ .
\end{equation}
Note that in the second (screening) term in (\ref{phi}) we
only keep the zero multipole $l=0$.
Also note that for  $R\rightarrow \infty$ the first and third terms
of Eq. (\ref{varphi1}) cancel each other. Therefore, we can use
$\int_{0}^{R}=\int_{0}^{\infty}-\int_{R}^{\infty}=-\int_{R}^{\infty}$
and present $\varphi ^{(1)}({\bf R})$ as
\begin{equation}
\label{phi1}
\varphi ^{(1)}({\bf R})=e{\bf R} \cdot \Big[
\int _{R}^{\infty}\left(
\frac{\langle {\bf r}\rangle}{ZR^{3}}-\frac{\bf r}{R^{3}}+
\frac{\bf r}{r^{3}}\right) \rho ({\bf r})d^{3}r \Big] \ .
\end{equation}
We see that $\varphi ^{(1)}({\bf R})=0$ for $R>R_{N}$ (nuclear radius)
 since $\rho ({\bf r})=0$ in this area.
We will see in the next section that this potential (\ref{phi1})
is related to the Schiff moment.


\section{Electron matrix elements of the $P,T$-odd electrostatic potential.}
\label{s:me}

All the electron orbitals for $l>1$ are extremely small inside the
nucleus. Therefore, we can limit our consideration to the matrix
elements between $s$ and $p$ Dirac orbitals. We will use the
following notations for the electron wavefunctions:
\begin{equation}
\label{psidef}
\psi ({\bf R})=
\left(
\begin{array}{c}
f(R)\Omega _{jlm} \\
-i(\mbox{\boldmath$\sigma$}\cdot {\bf n})g(R)\Omega _{jlm}
\end{array}
\right) \ ,
\end{equation}
where $\Omega _{jlm}$ is a spherical spinor, ${\bf n}={\bf R}/R$,
and f(R) and g(R) are radial functions
(see, e.g., \cite{khriplovich}).
Using $(\mbox{\boldmath$\sigma$}\cdot {\bf n})^{2}=1$, then we can write
the electron transition density as
\begin{eqnarray}
\label{rhodef}
&\rho _{sp}({\bf R})\equiv \psi _{s}^{\dagger}\psi _{p}=\Omega _{s}^{\dagger}
\Omega _{p}U_{sp}(R)&\\
\label{radwav}
&U_{sp}(R)=f_{s}(R)f_{p}(R)+g_{s}(R)g_{p}(R)=\sum_{k=1}^{\infty}b_{k}R^{k}\ .&
\end{eqnarray}
The expansion coefficients $b_{k}$ are calculated analytically and are
presented in Appendix \ref{append};
as is seen here, the summation is carried over
the odd powers of $k$.
Now we can find the matrix element of the electron-nucleus
interaction,
\begin{eqnarray}
\label{me}
\langle s|-e\varphi ^{(1)}({\bf R})|p\rangle &=&
-e^{2}\langle s|{\bf n}|p\rangle \cdot
\Big\{
\int _{0}^{\infty}
\Big[
\Big(
\frac{1}{Z}\langle {\bf r}\rangle-{\bf r}
\Big)
\int_{0}^{r}U_{sp}~dR+
\frac{{\bf r}}{r^{3}}\int_{0}^{r}U_{sp}R^{3}~dR\Big] \rho ~d^{3}r
\Big\} \nonumber\\
&=&-e^{2}\langle s|{\bf n}|p\rangle \cdot
\Big\{
\sum_{k=1}^{\infty}
\frac{b_{k}}{k+1}\Big[
\frac{1}{Z}\langle {\bf r}\rangle\langle r^{k+1}\rangle
-\frac{3}{k+4}\langle {\bf r}r^{k+1}\rangle
\Big] \Big\} \ ,
\end{eqnarray}
where $\langle s|{\bf n}|p\rangle \equiv
\int \Omega _{s}^{\dagger}{\bf n}\Omega_{p}d\phi \sin\theta d\theta$
and $\langle r^{n}\rangle \equiv \int \rho ({\bf r})r^{n}d^{3}r$.
Note that all vector values $\langle {\bf r}r^{n}\rangle$ are due to
the $P,T$-odd correction to the nuclear charge density $\rho$,
while $\frac{1}{Z}\langle r^{n}\rangle$ are
the usual $P,T$-even  moments of the charge density starting from the
mean-square radius $\frac{1}{Z}\langle r^{2}\rangle=r_{q}^{2}$ for
$k=1$.

In the non-relativistic case ($Z\alpha \rightarrow 0$) we have just
$b_{1}\neq 0$, and so
\begin{eqnarray}
\label{p}
\langle s|-e\varphi ^{(1)}|p\rangle &=&-\frac{e^{2}b_{1}}{2}
\langle s|{\bf n}|p\rangle \cdot
\Big[ \frac{1}{Z}\langle {\bf r}\rangle \langle r^{2}\rangle
-\frac{3}{5}\langle {\bf r}r^{2}\rangle \Big] \nonumber  \\
&=&4\pi e {\bf S}\cdot (\mbox{\boldmath$\nabla$}
\psi _{s}^{\dagger}\psi _{p})_{R\rightarrow 0} \ ,
\end{eqnarray}
where the Schiff moment ${\bf S}$ is defined as
\begin{equation}
\label{s}
{\bf S}=\frac{e}{10}\left[ \langle r^{2}{\bf r}\rangle
-\frac{5}{3Z}\langle r^{2}\rangle \langle {\bf r}\rangle \right]
=S~{\bf I}/I \ ,
\end{equation}
${\bf I}$ is the nuclear spin.
The expressions (\ref{p}), (\ref{s}) agree with the results of
Ref. \cite{FKS84}.

Therefore, Eq. (\ref{me}) gives us the possibility
of a consistent relativistic treatment of the atomic effects produced
by $P,T$-odd nuclear forces.
The nuclear and electronic problems can be separated in the following way.
The nuclear calculations can provide us with
the value of the {\it local dipole moment} (LDM)
\begin{equation}
\label{l}
{\bf L}=e\sum_{k=1}^{\infty}\frac{b_{k}}{b_{1}}
\frac{1}{(k+1)(k+4)}
\left[
\langle {\bf r}r^{k+1}\rangle
-\frac{k+4}{3Z}\langle {\bf r}\rangle \langle r^{k+1}\rangle
\right]
=L~{\bf I}/I
\end{equation}
which coincides with the Schiff moment ${\bf S}$ (\ref{s}) in the
non-relativistic limit ($Z\alpha \rightarrow 0$).
Note that this ``nuclear'' moment contains relativistic corrections
which arise from the {\it electron wavefunctions}
which are calculated analytically.
(It should further be noted that the corrections originating from the
$s$-$p_{1/2}$ and $s$-$p_{3/2}$ matrix elements are different.)
The electron matrix elements are then given by
\begin{equation}
\label{f}
\langle s|-e\varphi ^{(1)}|p\rangle =4\pi e{\bf L}\cdot
(\mbox{\boldmath$\nabla$}
\psi _{s}^{\dagger}\psi _{p})_{R\rightarrow 0}=
3e{\bf L}\cdot \langle s|{\bf n}|p\rangle
\left(
\frac{f_{s}f_{p}+g_{s}g_{p}}{R}
\right)
_{R\rightarrow 0} \ .
\end{equation}
These formulae (\ref{l},\ref{f}), in principle, solve the problem of the
consistent approach for the calculation of the interaction of the
relativistic electrons with the Schiff moment.
Note that to achieve $\sim 10\%$ accuracy
it is enough to keep in ${\bf L}$ just the first correction,
$b_{3}/b_{1}=-3/5~Z^{2}\alpha ^{2}/R_{N}^{2}$
for the $s$-$p_{1/2}$ matrix element and
$b_{3}/b_{1}=-9/20~ Z^{2}\alpha ^{2}/R_{N}^{2}$
for the $s$-$p_{3/2}$ matrix element (see Appendix \ref{append});
and at this level of accuracy ($\sim 10\%$) the values of the coefficients
$b_{3}/b_{1}$ (for $s$-$p_{1/2}$ and $s$-$p_{3/2}$) can be taken to be
the same.


\section{Local dipole moments induced by $P,T$-odd nuclear forces}
\label{s:ldm}

We can now calculate local dipole moments induced by $P,T$-odd nuclear forces.
We will begin in Section \ref{ss:extp} with the calculation of the
contribution of an external proton to the local dipole moment
of a spherical nucleus.
Because the best limit on the $P,T$-odd nucleon-nucleon interaction
has been extracted from ${^{199}{\rm Hg}}$ (which has an
external neutron, so only the core protons contribute
to the LDM) the result of Section \ref{ss:extp} is not
so interesting by itself. However, as will be explained in
Section \ref{ss:corep}, it provides us with a check
of the method for the calculation of the contribution of
core protons to the LDM.
Then in Section \ref{ss:coll} we calculate the collective LDM of an
octupole deformed nucleus.

The $P,T$-odd nucleon-nucleon interaction, to first-order in the
velocities $p/m$, can be presented as \cite{FKS84}
\begin{equation}
\label{wfull}
\hat{W}_{ab}=\frac{G}{\sqrt{2}}\frac{1}{2m}
\left( (\eta_{ab}\mbox{\boldmath$\sigma$}_{a}-
\eta_{ba}\mbox{\boldmath$\sigma$}_{b})\cdot \mbox{\boldmath$\nabla$}_{a}
\delta({\bf r}_{a}-{\bf r}_{b})+
\eta '_{ab}\left[ \mbox{\boldmath$\sigma$}_{a}
\times \mbox{\boldmath$\sigma$}_{b}\right]
\cdot \left\{
({\bf p}_{a}-{\bf p}_{b}),\delta({\bf r}_{a}-{\bf r}_{b})
\right\}
\right),
\end{equation}
where $\{\ ,\ \}$ is an anticommutator, $G$ is the
Fermi constant of the weak interaction, $m$ is the nucleon mass,
and \mbox{\boldmath$\sigma$}, ${\bf r}$, and ${\bf p}$ are the spins,
coordinates, and momenta of the nucleons $a$ and $b$.
The dimensionless constants $\eta _{ab}$ and $\eta '_{ab}$
characterize the strength of the
$P,T$-odd nuclear potential
(experiments on EDMs are aimed to measure these constants).


\subsection{Nuclear LDM produced by external proton}
\label{ss:extp}

In this section we calculate the LDM arising due to an external proton.
We are therefore interested in the $P,T$-odd interaction of
the external proton with the core nucleons.
We can average the two-particle interaction (\ref{wfull})
over the core nucleons to obtain the effective single-particle $P,T$-odd
interaction between the proton and core \cite{FKS84},
\begin{equation}
\label{w}
\hat{W}=\frac{G}{\sqrt{2}}\frac{\eta}{2m}
\mbox{\boldmath$\sigma$}\cdot \mbox{\boldmath$\nabla$}
\rho _{A}({\bf r}) \ .
\end{equation}
Here it has been assumed that the proton and neutron densities
are proportional to the total nuclear density $\rho _{A}({\bf r})$;
the dimensionless constant $\eta =\frac{Z}{A}\eta _{pp} +
\frac{N}{A}\eta _{pn}$.
Notice that there is only one surviving term from the $P,T$-odd
nucleon-nucleon interaction (\ref{wfull}); this is because all
other terms contain the spin of the internal nucleons for which
$\langle \mbox{\boldmath$\sigma$}\rangle =0$.
The shape of the nuclear density $\rho _{A}$ and the strong potential
$U$ are known to be similar; we therefore take
$\rho _{A}({\bf r})=\frac{\rho _{A}(0)}{U(0)}U({\bf r})$.
Then we can rewrite Eq. (\ref{w}) in the following form:
\begin{equation}
\hat{W}=\xi \mbox{\boldmath$\sigma$}\cdot \mbox{\boldmath$\nabla$}U \ , \qquad
\xi=\eta \frac{G}{2\sqrt{2}m}\frac{\rho _{A}(0)}{U(0)}=
-2\times 10^{-8}\eta~{\rm fm} \ .
\end{equation}
Now it is easy to find the solution of the Schr{\"o}dinger equation
including the interaction $\hat{W}$ \cite{FKS84}:
\begin{eqnarray}
&(\hat{H}+\hat{W})\tilde {\psi}=E\tilde{\psi} \ ,  & \nonumber \\
&\tilde{\psi}=(1+\xi \mbox{\boldmath$\sigma$}
\cdot \mbox{\boldmath$\nabla$})\psi \ , &
\end{eqnarray}
where $\psi$ is the unperturbed solution ($\hat{H}\psi =E\psi$).
The valence proton density is then equal to
\begin{equation}
\label{density}
\rho =\tilde{\psi}^{\dagger}\tilde{\psi}=
\psi ^{\dagger}\psi+ \xi \mbox{\boldmath$\nabla$}\cdot
(\psi ^{\dagger}\mbox{\boldmath$\sigma$}\psi)\ .
\end{equation}
The second term gives the $P,T$-odd part of the density which
generates a Schiff moment $S$ \cite{FKS84},
\begin{equation}
\label{schiff84}
S=-\frac{e\xi}{10}
\Big[ \Big( t_{I}+\frac{1}{I+1}\Big) r_{\rm ex}^{2}-
\frac{5}{3}t_{I}r_{q}^{2}\Big] \ ,
\end{equation}
where we denote $r_{\rm ex}^{2}$ as the mean-square radius of the
external nucleon (in this case that of the proton),
$r_{q}^{2}$ is the mean-square nuclear charge radius,
$\langle \mbox{\boldmath$\sigma$}\rangle=t_{I}\frac{\bf I}{I}$, and
\begin{equation}
t_{I}=
\left\{
\begin{array}{ll}
1 \qquad & I=l+1/2 \\
-\frac{I}{I+1} \qquad & I=l-1/2
\end{array}
\right.
\end{equation}
where $I$ and $l$ are the total and orbital angular momentum of the proton,
respectively.
It should be noted that for the Schiff moment the recoil effect
(the motion of the nuclear core around the center-of-mass)
disappears due to the cancellation of its contributions to the
first and second (screening) terms in Eq. (\ref{phi}) \cite{FKS84}.

To calculate the local dipole moment, it is enough to substitute
the external proton density (\ref{density}) into the expression for the LDM
(\ref{l}) and perform integration using integration by parts. The result is
\begin{equation}
\label{ldmextp}
L=-\sum_{k=1}^{\infty}\frac{b_{k}}{b_{1}}
\frac{e\xi}{(k+1)(k+4)}
\left[
\left( t_{I}+\frac{k+1}{2(I+1)} \right)
r^{k+1} _{\rm ex}
- \frac{k+4}{3}t_{I} r_{q}^{k+1}
\right] \ .
\end{equation}
(Notice that for a proton in the state $s_{1/2}$,
the LDM is reduced to the difference of two approximately equal terms
($r_{\rm ex}^{k}-r_{q}^{k}$) for all $k=2,4,...$.
This makes an analytical calculation hopeless when
trying to estimate the LDM of a nucleus with an external proton in state
$s_{1/2}$, as is the case for ${^{203,205}{\rm Tl}}$.)


\subsection{Nuclear LDM produced by core protons. Mercury moments.}
\label{ss:corep}

In nuclei like ${^{199}{\rm Hg}}$ and ${^{129}{\rm Xe}}$
the external nucleon is a neutron.
It does not contribute to the Schiff moment directly.
In Ref. \cite{FKS86} it was shown that virtual
excitations of the core nucleons caused by a $P,T$-odd interaction
with the external nucleon produce a Schiff moment which is comparable
to that produced by an external proton.
The actual calculation of the Schiff
moments in \cite{FKS86} was carried out numerically.
In this section we perform a simple analytical calculation which
allows us to estimate the contribution of the relativistic
corrections $\sim Z^{2}{\alpha}^{2}$ to the Schiff moment.
Here we follow an approach which was used in \cite{FKS86} to estimate
the contribution of the giant dipole resonance to the nuclear EDM.

The expression for the local dipole moment ${\bf L}$ induced
in the nuclear state $|0\rangle $ by the
$P,T$-odd interaction $\hat{W}_{ab}$ (\ref{wfull})
between the nucleons $a$ and $b$ is
\begin{eqnarray}
\label{first}
{\bf L}&=&\sum_{n}\frac{\langle 0|\hat{\bf L}|n\rangle
\langle n|\hat{W}_{ab}|0\rangle + \langle 0|\hat{W}_{ab}|n\rangle
\langle n|\hat{\bf L}|0\rangle}{E_{0}-E_{n}}\\
\label{second}
&=&\sum_{n}\frac{\langle 0|[\hat{H},\hat{\bf L}]|n\rangle
\langle n|\hat{W}_{ab}|0\rangle - \langle 0|\hat{W}_{ab}|n\rangle
\langle n|[\hat{H},\hat{\bf L}]|0\rangle}{(E_{0}-E_{n})^{2}} \ .
\end{eqnarray}
Here $\hat{H}$ is the Hamiltonian, and $[\hat{H},\hat{\bf L}]$
is a commutator; the LDM operator $\hat{\bf L}$ is
defined from Eq. (\ref{l}) as ${\bf L}=\langle \hat{\bf L}\rangle$.
Now we assume that the transition strength
in the sum over intermediate states $|n\rangle$ is concentrated
around the excitation energy $\omega _{r}$, and replace
$(E_{0}-E_{n})^{2}$ by $\omega _{r}^{2}$.
(Note that the replacement of $(E_{0}-E_{n})$ by $\omega _{r}$
in Eq. (\ref{first}) gives an incorrect result since in
single-particle language there are transitions with
$E_{0}-E_{n}=\omega_{r}$ and $E_{0}-E_{n}=-\omega _{r}$.)
Use of closure, $\sum_{n}|n\rangle \langle n|=1$, gives
\begin{equation}
{\bf L}=\frac{1}{\omega _{r}^{2}}
\langle 0|\left[ [ \hat{H},\hat{\bf L}] ,\hat{W}_{ab}\right] |0\rangle \ .
\end{equation}
To calculate the commutator we assume that the motion of each nucleon
in the nucleus can be described by the Hamiltonian
$\hat{H}=\frac{\hat{p}^{2}}{2m}+V(r)$.
The contribution to the LDM from a single proton is then
\begin{equation}
\label{schiffres}
{\bf L}=-\frac{1}{m\omega _{r}^{2}}\langle 0|
(\nabla _{\alpha}\hat{\bf L})(\nabla _{\alpha}\hat{W}_{ab})|0\rangle .
\end{equation}

As a check of the validity of this approach for the calculation of the
core contribution, we use this formula (\ref{schiffres}) to calculate
both the contribution of the external proton (which we can compare
with Eq. (\ref{ldmextp})) and the contribution of the core protons.

\subsubsection{External proton contribution}

For the external proton we can just substitute the effective potential
$\hat{W}$ (\ref{w}) into expression (\ref{schiffres}),
\begin{equation}
{\bf L}=-\frac{G}{\sqrt{2}}\frac{1}{2m^{2}\omega _{r}^{2}}
\eta
\langle 0|(\nabla _{\alpha}\hat{\bf L})(\nabla _{\alpha}
(\mbox{\boldmath$\sigma$}
\cdot \mbox{\boldmath$\nabla$})\rho _{A})|0\rangle ,
\end{equation}
where here $|0\rangle$ corresponds to the state of the
external proton.
If we consider the nuclear density $\rho _{A}$ to be proportional to
the nuclear potential $U$, and use the oscillator model to approximate
the potential so that
\begin{equation}
\label{densityho}
\rho _{A} (r)=\frac{\rho _{A}(0)}{U(0)}\frac{1}{2}m\omega ^{2}r^{2} \ ,
\end{equation}
then the LDM is reduced to
\begin{equation}
L=-\frac{\omega ^{2}}{\omega _{r}^{2}}
\sum_{k=1}^{\infty}\frac{b_{k}}{b_{1}}
\frac{e\xi}{(k+1)(k+4)}
\left[
\left( t_{I}+\frac{k+1}{2(I+1)} \right)
r^{k+1} _{\rm ex}
- \frac{k+4}{3}t_{I} r_{q}^{k+1}
\right] \ .
\end{equation}
Because here we are considering the LDM produced by a single proton,
we can set $\omega _{r}=\omega$,
so the factor $(\omega ^{2}/\omega _{r}^{2})=1$.
Therefore, using the resonance method we can reproduce Eq. (\ref{ldmextp}).

\subsubsection{Core contribution}

In this section we use the ``Schiff resonance'' formalism
to estimate the local dipole moment produced by the core protons.
Again we start from Eq. (\ref{schiffres}), where in this case
the derivatives are with respect to the internal proton coordinates.
Assuming that the proton density is proportional to the total nuclear
density $\rho _{A}$, we obtain
\begin{equation}
{\bf L}=\frac{G}{2\sqrt{2}m}\frac{1}{m\omega _{r}^{2}}\eta
\langle 0|\mbox{\boldmath$\sigma$}\cdot \mbox{\boldmath$\nabla$}
\Big[
\nabla _{\alpha}\Big( \rho _{A} \nabla _{\alpha}\hat{\bf L}\Big)
\Big] |0\rangle \ ,
\end{equation}
where $|0\rangle$ is the state of the external nucleon.
Because here we are considering the $P,T$-odd interaction
$\hat{W}_{ap}$ (\ref{wfull}) between the external nucleon
(proton or neutron, $a$) and the core protons, the
$P,T$-odd dimensionless constant $\eta =(Z/A)\eta _{ap}$.
Using (\ref{densityho}) to approximate the nuclear density,
the LDM becomes
\begin{equation}
\label{ldmcore}
L=\frac{\omega ^{2}}{\omega _{r}^{2}}
\sum_{k=1}^{\infty} \frac{b_{k}}{b_{1}}
\frac{e\xi}{(k+1)(k+4)}
\left[
\frac{k^{2}+7k+8}{2}
\left(
t_{I}+
\frac{k+1}{2(I+1)}
\right)
r_{\rm ex}^{k+1}
-\frac{k+4}{3}t_{I}r_{q}^{k+1}
\right] \ .
\end{equation}
Before we consider the size of the relativistic corrections,
let us check that this result gives us a reasonable
value for the Schiff moment ($Z\alpha \rightarrow 0$).
In the approximation of a uniform and spherical charge distribution,
$r_{q}^{2}=(3/5)R_{N}^{2}$, assuming that $r_{\rm ex}^{2}=r_{q}^{2}$,
and setting the resonance frequency for core protons
$\omega _{r}=2\omega$ (frequency of giant resonance),
we obtain for the Schiff moment of ${^{199}{\rm Hg}}$
(which has an external neutron in the state $p_{1/2}$)
\begin{equation}
\label{schiffhg}
S_{\rm Hg}\approx -1.6\times 10^{-8}\eta _{np}e~{\rm fm}^{3} \ .
\end{equation}
From a numerical calculation of the Schiff moment for ${^{199}{\rm Hg}}$
performed in Ref. \cite{FKS86} the result
$S=-1.4\times 10^{-8}\eta _{np}e~{\rm fm}^{3}$
was obtained.
This value agrees with our analytical estimate (\ref{schiffhg}).
Therefore it seems that the resonance method can be used for Hg
to give a crude estimate of the size of the relativistic corrections
to the Schiff moment.

To estimate the size of the corrections, we use the approximation of a
uniform and spherical charge distribution, $r_{q}^{n}=(3/(n+3))R_{N}^{n}$,
and we assume that $r_{\rm ex}^{n}=r_{q}^{n}$, for $n=2,4,...$.
Substituting the coefficients from Appendix \ref{append} into the
expression for the LDM (\ref{ldmcore}),
we see that the first correction ($k=3$) to the Schiff moment
for Hg is about $25\%$.
(Note that the second correction ($k=5$) amounts to less than $10\%$.)
Therefore the relativistic corrections to the Schiff moment for
Hg (and we expect for other spherical nuclei) are not very large;
for Hg we have $L\approx 0.75 S$.


\subsection{Collective LDM of an octupole deformed nucleus}
\label{ss:coll}

Nuclei with octupole deformation have enhanced collective Schiff moments which
may be up to $1000$ times larger than the Schiff moments of spherical nuclei
\cite{AFS,SAF}. In Ref. \cite{hayes} it was pointed out that the soft octupole
vibration mode produces an enhancement similar to that of the static
octupole deformation. This makes heavy atoms containing nuclei with
collective Schiff moments attractive for future experiments searching
for $T$-violation.

The mechanism generating the collective Schiff moment is the following
\cite{AFS,SAF}.
In the ``frozen'' body frame the collective Schiff moment
$S_{\rm intr}$ can exist without any $P,T$-violation.
However, the nucleus rotates, and this makes the expectation value of
the Schiff moment vanish in the laboratory frame
if there is no $P,T$-violation.
(This is because the intrinsic Schiff moment is directed along the
nuclear axis, ${\bf S}_{\rm intr}=S_{\rm intr}{\bf n}$,
and in the laboratory frame the only possible correlation
$\langle {\bf n}\rangle \propto {\bf I}$ violates parity and time reversal.)
The $P,T$-odd nuclear forces mix rotational states of
opposite parity and create an average orientation of the nuclear axis
${\bf n}$ along the nuclear spin ${\bf I}$,
\begin{equation}
\label{nz}
\langle n_{z}\rangle =2\alpha \frac{KM}{I(I+1)} \ ,
\end{equation}
where
\begin{equation}
\alpha =\frac{\langle \psi _{-}|\hat{W}|\psi _{+}\rangle}{E_{+}-E_{-}}
\end{equation}
is the mixing coefficient of the opposite parity states,
$K=|{\bf I}\cdot {\bf n}|$ is the absolute value of the
projection of the nuclear spin ${\bf I}$ on the nuclear axis,
$M=I_{z}$, and $\hat{W}$ is the effective single-particle potential
(\ref{w}).
The Schiff moment in the laboratory frame is
\begin{equation}
S_{z}=S_{\rm intr}\langle n_{z}\rangle =S_{\rm intr}\frac{2\alpha KM}{I(I+1)}
\ .
\end{equation}

In the ``frozen'' body frame the surface of an axially
symmetric deformed nucleus is described by the following expression
\begin{equation}
\label{defsurf}
R(\theta)=R_{N}\big( 1+\sum_{l=1}\beta_{l}Y_{l0}(\theta)\big) \ .
\end{equation}
To keep the center-of-mass at $r=0$ we have to fix $\beta _{1}$
\cite{bm}:
\begin{equation}
\label{beta1}
\beta_{1}=-3\sqrt{\frac{3}{4\pi}}\sum_{l=2}\frac{(l+1)\beta_{l}\beta_{l+1}}
{\sqrt{(2l+1)(2l+3)}} \ .
\end{equation}
We assume that the distributions of the protons and neutrons are the same,
so the electric dipole moment $e\langle {\bf r}\rangle=0$
(since the center-of-mass of the
charge distribution coincides with the center-of-mass)
and hence there is no screening contribution to the Schiff moment.
We also assume constant density for $R<R(\theta)$.
The intrinsic Schiff moment $S_{\rm intr}$ is then \cite{AFS,SAF}
\begin{equation}
\label{intrschiff}
S_{\rm intr}=eZR_{N}^{3}\frac{3}{20\pi}\sum_{l=2}
\frac{(l+1)\beta _{l}\beta _{l+1}}{\sqrt{(2l+1)(2l+3)}}
\approx eZR_{N}^{3}\frac{9\beta _{2}\beta_{3}}{20\pi \sqrt{35}} \ ,
\end{equation}
where the major contribution comes from $\beta_{2}\beta_{3}$, the
product of the quadrupole $\beta_{2}$ and octupole $\beta _{3}$
deformations. For $\beta _{2}\sim \beta_{3}\sim 0.1$ and
$Z=88$ (Ra) we obtain $S_{\rm intr}\sim 10 ~e{\rm fm}^{3}$.
The estimate of the Schiff moment in the laboratory frame gives~\cite{SAF}
\begin{equation}
\label{snum}
S\sim \alpha S_{\rm intr}\sim 0.05 ~e \beta_{2}\beta_{3}^{2}ZA^{2/3}\eta
r_{0}^{3}\frac{\rm eV}{E_{+}-E_{-}}\sim 700 \times
10^{-8}\eta e~{\rm fm}^{3} \ ,
\end{equation}
where $r_{0}\approx 1.2~{\rm fm}$ is the internucleon distance,
$E_{+}-E_{-}\sim 50~{\rm keV}$. This estimate (\ref{snum}) is about
$500$ times larger than the Schiff moment of a spherical nucleus
like Hg. Note that $S$ in Eq. (\ref{snum}) is proportional
to the squared octupole deformation parameter $\beta_{3}^{2}$.
According to \cite{hayes}, in nuclei with a soft octupole vibration
mode $\langle \beta _{3}^{2}\rangle \sim (0.1)^{2}$, i.e., this is the
same as in nuclei with static octupole deformation. This means that
a number of heavy nuclei can have large collective Schiff moments.

With no screening term, it is easy to calculate the collective LDM.
Use of Eq.~(\ref{l}) gives
\begin{equation}
\label{ldmdef}
L=S\left( 1+\sum_{k=3}\frac{5b_{k}}{(k+4)b_{1}}R_{N}^{k-1}\right)
\approx S(1-0.35Z^{2}{\alpha}^{2})\ .
\end{equation}
As with spherical nuclei, we see that the correction to the Schiff
moment for collective nuclei is not very large
(for Ra, Pu this correction $\sim 15\%$).


\section{$P,T$-odd part of the nuclear electric field
(Schiff field)}
\label{s:efield}

In this section we calculate the actual distribution of the
$P,T$-odd component of the electrostatic potential
$\varphi ({\bf R})$ inside the nucleus
(arising from the $P,T$-odd nucleon-nucleon interaction) for two models:
for an external proton in a spherical nucleus and
for a collective Schiff moment which appears
in a nucleus with octupole deformation.
It is found that in the collective case the electric field
is constant and directed along the nuclear spin.
This field distribution is also approximately correct
in the spherical case when the external proton is in state $s_{1/2}$;
this is also true without the $P,T$-odd interaction but when the
external nucleon (proton or neutron) possesses
an intrinsic EDM and is in state $s_{1/2}$.

\subsection{$P,T$-odd electric field produced by a valence proton}

To calculate the $P,T$-odd part of the electrostatic potential
$\varphi _{T}$ produced by the external proton, we substitute the
$P,T$-odd perturbed external proton density (\ref{density})
into (\ref{phi}) and integrate by parts,
\begin{equation}
\varphi _{T}({\bf R})=
e\xi \mbox{\boldmath$\nabla$}\cdot \Big[
\int \frac{\mbox{\boldmath$\rho$}_{\sigma}}
{|{\bf R}-{\bf r}|}d^{3}r \Big] - \frac{1}{Z}e\xi
\langle \mbox{\boldmath$\sigma$}\rangle \cdot \mbox{\boldmath$\nabla$}
\int \frac{\rho}{|{\bf R}-{\bf r}|}d^{3}r \ ,
\end{equation}
where
$\mbox{\boldmath$\rho$} _{\sigma}=\psi ^{\dagger}\mbox{\boldmath$\sigma$}\psi$
is the spin density,
$\langle \mbox{\boldmath$\sigma$}\rangle =t_{I}\frac{\bf I}{I}$.
Note the similarity of this expression and that for a $P,T$-odd potential
produced by an external proton electric dipole moment $d_{p}$
(see, e.g., \cite{schiff,khriplovich}).
In the latter case one should only replace $e\xi$ by $d_{p}$
(or by $d_{n}$ in the case of Hg or Xe).
Note, however, that generally speaking
$\mbox{\boldmath$\rho$}_{\sigma}=\psi ^{\dagger}\mbox{\boldmath$\sigma$}\psi
\neq \langle \mbox{\boldmath$\sigma$}\rangle \psi ^{\dagger}\psi$
(this assumption was used in \cite{schiff}),
i.e. the direction of the external nucleon EDM depends on the coordinate
${\bf r}$. The separation of the spin and the coordinate variables is
possible in the case of $I=l+1/2$. Taking $I=I_{z}$ we obtain
$\mbox{\boldmath$\rho$}_{\sigma}=
\frac{\bf I}{I}\rho _{M}$, where $\rho _{M}$ is the
density of the valence proton. If we now assume that this density
is constant within the sphere of the radius $R_{M}$,
$\rho _{M}(r)=\frac{3}{4\pi R_{M}^{3}}$, $r<R_{M}$, and
$\rho _{M}(r)=0$ for $r>R_{M}$, and, similarly, the nuclear charge density
$\rho _{q}(r)=\frac{3Z}{4\pi R_{N}^{3}}$, $r<R_{N}$, and
$\rho _{q}(r)=0$ for $r>R_{N}$, then we obtain for the dipole term
(and for $R_{N}<R_{M}$):
\begin{equation}
\label{phiconst}
\varphi ^{(1)}({\bf R})=-e\xi {\bf R}\cdot \frac{\bf I}{I}
\left\{
\begin{array}{ll}
(\frac{1}{R_{M}^{3}}-\frac{1}{R_{N}^{3}}) \qquad & R<R_{N} \\
(\frac{1}{R_{M}^{3}}-\frac{1}{R^{3}}) \qquad & R_{N}<R<R_{M}\\
0 \qquad & R>R_{M} \ .
\end{array}
\right.
\end{equation}
Thus, the $P,T$-odd part of the electrostatic potential is
$\varphi ^{(1)}({\bf R})\propto R\cos \theta$ inside the nucleus.
This gives us a very simple picture for the $P,T$-odd electric field
(Schiff field): $E_{z}=-\frac{\partial \varphi}{\partial z}\propto I$
inside the nucleus. Thus, the Schiff moment gives a constant
electric field along the nuclear spin, ${\bf E}\propto {\bf I}$,
and this field vanishes within the nuclear ``skin''
(see Fig. \ref{fig:schiff}).

We can easily establish a relation between the $P,T$-odd electrostatic
potential inside the nucleus $\varphi ^{(1)}\propto R\cos\theta$
and the Schiff moment ${\bf S}$.
Comparing Eq. (\ref{phiconst}) with
Eq. (\ref{schiff84}) (with $t_{I}=1$, $I=1/2$)
we obtain
\begin{equation}
\label{ptoddpots}
\varphi ^{(1)}({\bf R})= -\frac{15{\bf S}\cdot{\bf R}}{R_{N}^{5}}n(R-R_{N}) ,
\end{equation}
where $n(R-R_{N})$ is a smoothed step-function $\Theta(R_{N}-R)$,
that is $n(R-R_{N})\approx \Theta (R_{N}-R)$;
$n(R-R_{N})=1$ for $R<R_{N}$ and
$n(R-R_{N})=0$ for $R>R_{N}+\delta$,
where $\delta =R_{M}-R_{N}<<R_{N}$.
It gives the natural generalization
of the Schiff moment potential (\ref{olddef}) for the case of a
finite-size nucleus.
Of course, in the general case, the radial function $n(R-R_{N})$ in the
first harmonic of the $P,T$-odd potential
(\ref{ptoddpots}) is more complicated
(this gives some ``wiggling'' of the electric field inside
the nucleus).

\subsection{$P,T$-odd electric field produced by a collective Schiff moment}

Now we wish to calculate the electrostatic potential $\varphi ^{(1)}$
arising due to a collective Schiff moment in a nucleus
with octupole deformation.
We use Eqs. (\ref{defsurf},\ref{beta1}) and assume
that the distributions of the protons and neutrons are the same
(therefore $e\langle {\bf r}\rangle =0$, and so there is no screening term)
and that the density for $R<R(\theta)$ is constant.
Calculating the integral in Eq. (\ref{phi1}) for $R<{\rm min~}R(\theta)$
gives
\begin{equation}
\varphi ^{(1)}({\bf R})=-\frac{9Ze}{4\pi R_{N}^{2}}{\bf R}\cdot
\frac{\bf I}{I}
\sum_{l=2}\frac{(l+1)\beta_{l}\beta_{l+1}}{\sqrt{(2l+1)(2l+3)}} \ .
\end{equation}
In the laboratory frame the result differs by an extra factor
$\langle n_{z}\rangle$ (\ref{nz}).
Using Eq. (\ref{intrschiff}) we can present the final result for
$\varphi ^{(1)}$ as
\begin{equation}
\label{phi1fins}
\varphi ^{(1)}({\bf R})=-\frac{15{\bf S}\cdot {\bf R}}{R_{N}^{5}}
n(R-R_{N})\ ,
\end{equation}
where $n(R-R_{N})=1$ for $R<{\rm min~}R(\theta)$,
and $n(R-R_{N})=0$ for $R>{\rm max~}R(\theta)$.
This result is similar to Eq. (\ref{ptoddpots}).
Thus, the collective Schiff moment produces a constant electric field along
the nuclear spin inside the nucleus and zero field outside
(Fig.~\ref{fig:schiff}).
In this case the width of the transition area of the nuclear
surface $\sim \beta_{l}R_{N}\sim 0.2R_{N}$.

The expression (\ref{ptoddpots},\ref{phi1fins}) contains the fifth
power of the nuclear radius $R_{N}$ which is not a very well-defined
property in the case of finite nuclear thickness.
We can avoid this ambiguity by using a ``normalized'' expression
which is exact in the limit $Z\alpha <<1$ when $\rho _{sp}\sim R$
(compare with Eq. \ref{eq:nonrelS}):
\begin{equation}
\varphi ^{(1)}({\bf R})=
-\frac{3{\bf S}\cdot {\bf R}}{B}n(R-R_{N}) \quad , \qquad
B=\int n(R-R_{N})R^{4}dR \approx R_{N}^{5}/5 \ .
\end{equation}


\section{The atomic EDM induced in plutonium}
\label{s:atomicedm}

In Ref. \cite{hayes} it was shown that ${^{239}{\rm Pu}}$
has a large vibrational Schiff moment and it was pointed out that it is
a good candidate for experiments searching for $P,T$-odd effects:
it has a ground state nuclear spin $I=1/2$ and it has a long half-life.
The ground-state electron angular momentum is $J=0$, and so the atomic EDM
is sensitive to the nuclear Schiff moment;
however, it corresponds to a complex electronic configuration,
$5f^{6}7s^{2}$.

In this section we perform a simple analytical estimate of the size of the
atomic electric dipole moment induced by the nuclear Schiff moment for
${^{239}{\rm Pu}}$.
The contribution of electrons from the $f$-shell is small.
Therefore, in a simplistic model we can consider ${^{239}{\rm Pu}}$
to be an electronic analog of ${^{199}{\rm Hg}}$.
We can then exploit the $Z$-dependence of the induced atomic EDM,
as was done, e.g., in Ref. \cite{SAF},
to estimate the EDM of ${^{239}{\rm Pu}}$ from the calculation
of the atomic EDM of ${^{199}{\rm Hg}}$.
The arguments for this simple estimate follow.

We see from (\ref{atomedmform}) that there are three factors
contributing to the atomic EDM:
the electric dipole transition amplitudes, the energy denominators,
and the Schiff matrix elements.
The first two factors are sensitive to the wavefunctions at
large distances, and these are similar for analogous atoms.
The matrix element of the Schiff moment is determined from distances
close to the nucleus, and therefore the significant contributions
come from the matrix elements of
$s_{1/2}$ and $p_{1/2}$ as well as $s_{1/2}$ and $p_{3/2}$.
These matrix elements strongly depend on the nuclear charge:
they are proportional to $SZ^{2}R_{1/2}$ and
$SZ^{2}R_{3/2}$, respectively, where the relativistic enhancement factors
$R_{1/2}$ and $R_{3/2}$ are given by \cite{FKS84}
\begin{eqnarray}
R_{1/2}&=&
\frac{4\gamma _{1/2}}{[\Gamma (2\gamma _{1/2}+1)]^{2}}
{\Big(
\frac{2ZR_{N}}{a_{B}}\Big) ^{2\gamma _{1/2}-2}} \ ,
\nonumber \\
R_{3/2}&=&
\frac{48}{\Gamma (2\gamma _{1/2}+1) \Gamma (2\gamma _{3/2}+1)}
{\Big(
\frac{2ZR_{N}}{a_{B}}\Big) ^{\gamma _{1/2}+\gamma _{3/2}-3}} \ ,
\end{eqnarray}
where
$\gamma _{j}=[(j+\frac{1}{2})^{2}-(Z\alpha )^{2}]^{\frac{1}{2}}$ and
$a_{B}$ is the Bohr radius.
Because there are twice as many $p_{3/2}$ states as $p_{1/2}$ states,
we use a linear combination of $R_{1/2}$ and $R_{3/2}$ in the calculations,
$R_{sp}=(R_{1/2}+2R_{3/2})/3$.

Our estimate for the atomic EDM induced in ${^{239}{\rm Pu}}$ due to
its nuclear Schiff moment can therefore be expressed in terms of the
Schiff moment and atomic EDM of ${^{199}{\rm Hg}}$, for which
calculations have been performed,
\begin{equation}
d_{\rm atom}({\rm Pu})
=d_{\rm atom}({\rm Hg})
\frac{(SZ^{2}R_{sp})_{\rm Pu}}{(SZ^{2}R_{sp})_{\rm Hg}} \ .
\end{equation}
We use the results of Ref. \cite{FKS86} for the value of the Schiff moment
of Hg, $S_{\rm Hg}=-1.4\times 10^{-8} \eta _{np}e~{\rm fm}^{3}$,
and the atomic EDM it induces,
$d_{\rm atom}({\rm Hg})=5.6\times 10^{-25} \eta _{np}e~{\rm cm}$.
The atomic structure ratio
$(Z^{2}R_{sp})_{\rm Pu}/(Z^{2}R_{sp})_{\rm Hg}=2.6$.
The atomic EDM for ${^{239}{\rm Pu}}$ in terms of its Schiff moment is then
\begin{equation}
d_{\rm atom}({\rm Pu})=-1\times 10^{-16}
(S_{\rm Pu}/(e~{\rm fm ^{3}}))e~{\rm cm} \ .
\end{equation}
If we take the value for the Schiff moment of Pu from Ref. \cite{hayes},
$S_{\rm Pu}=400\times 10^{-8} \eta _{n}e~{\rm fm}^{3}$,
then the atomic EDM is
$d_{\rm atom}({\rm Pu})=-4\times 10^{-22} \eta _{n}e~{\rm cm}$,
which is $\approx 700$ times larger than the atomic EDM
induced in ${^{199}{\rm Hg}}$.

A numerical calculation of the atomic EDMs induced in
Hg, Xe, Rn, Ra, and Pu is underway.

\acknowledgments

We are grateful to V.A. Dzuba for useful discussions.
This work was supported by the Australian Research Council.

\appendix

\section{Electron transition density for $s$-$p_{1/2}$ and $s$-$p_{3/2}$}
\label{append}

To calculate the electron wavefunctions inside the nucleus $R\leq R_{N}$
we assume that the nuclear charge is uniformly distributed about a sphere.
This charge distribution corresponds to the harmonic-oscillator potential
\begin{equation}
V=-\frac{Z\alpha}{R_{N}}\left( \frac{3}{2}-
\frac{1}{2}x ^{2}\right) \ ,
\end{equation}
where we have set $x\equiv R/R_{N}$.
By solving the radial Dirac equations for an electron in states $s_{1/2}$,
$p_{1/2}$, and $p_{3/2}$ moving in this potential we obtain
the radial wavefunctions (see Eq. \ref{psidef} for definition)
for $s$:
\begin{eqnarray}
f_{s}(R)&=&f_{s}(0)
\Big[
1-\frac{3}{8}Z^{2}{\alpha ^{2}}x^{2}\Big( 1-\frac{4}{15}x^{2}+
\frac{1}{45}x^{4}-\frac{9}{80}Z^{2}{\alpha ^{2}}x^{2}
+ ...
\Big) \nonumber \\
&&-\frac{1}{2}R_{N}mZ\alpha x^{2}\Big(
1-\frac{1}{10}x^{2}-\frac{9}{40}Z^{2}{\alpha ^{2}}x^{2}
+ ...
\Big)
\Big] \nonumber \\
g_{s}(R)&=&f_{s}(0)
\Big[
-\frac{1}{2}Z\alpha x
\Big(
1-\frac{1}{5}x^{2}-\frac{9}{40}Z^{2}{\alpha ^{2}}x^{2}
+ ...
\Big) \nonumber \\
&&+\frac{3}{20}R_{N}mZ^{2}{\alpha ^{2}}x^{3}\Big(
1-\frac{13}{42}x^{2}+\frac{1}{54}x^{4}-\frac{9}{56}Z^{2}{\alpha ^{2}}x^{2}
+...
\Big)
\Big] \ ; \nonumber
\end{eqnarray}
for $p_{1/2}$:
\begin{eqnarray}
f_{p_{1/2}}(R)&=& g_{p_{1/2}}(0)
\Big[
\frac{1}{2}Z\alpha x
\Big(
1-\frac{1}{5}x^{2}-\frac{9}{40}Z^{2}{\alpha ^{2}}x^{2}
+...
\Big) \nonumber \\
&&+\frac{2}{3}R_{N}mx
\Big\{
1-\frac{9}{20}Z^{2}{\alpha ^{2}}x^{2}
\Big(
1-\frac{23}{84}x^{2}+\frac{7}{324}x^{4}-\frac{27}{224}Z^{2}{\alpha ^{2}}x^{2}
+...
\Big)
\Big\}
\Big] \nonumber \\
g_{p_{1/2}}(R)&=& g_{p_{1/2}}(0)
\Big[
1-\frac{3}{8}Z^{2}{\alpha ^{2}}x^{2}\Big( 1-\frac{4}{15}x^{2}+
\frac{1}{45}x^{4}-\frac{9}{80}Z^{2}{\alpha ^{2}}x^{2}
+...
\Big) \nonumber \\
&&-\frac{1}{2}R_{N}mZ\alpha x^{2}
\Big(
1-\frac{1}{6}x^{2} -\frac{9}{40}Z^{2}{\alpha ^{2}}x^{2}
+...
\Big)
\Big] \ ; \nonumber
\end{eqnarray}
and for $p_{3/2}$:
\begin{eqnarray}
&& f_{p_{3/2}}(R)=(f_{p_{3/2}}/x)_{R=0}x
\Big[
1-\frac{9}{40}Z^{2}{\alpha ^{2}}x^{2}
\Big(
1-\frac{2}{7}x^{2}+\frac{5}{189}x^{4}-\frac{9}{112}Z^{2}{\alpha ^{2}}x^{2}
+ ...
\Big)
\Big] \nonumber \\
&& g_{p_{3/2}}(R)=-Z\alpha (f_{p_{3/2}}/x)_{R=0}x^{2}
\Big[
\frac{3}{10}-\frac{1}{14}x^{2}-\frac{27}{560}Z^{2}{\alpha ^{2}}x^{2}
+ ...
\Big] \ . \nonumber
\end{eqnarray}
Here $f_{s}(0)$, $g_{p_{1/2}}(0)$, and $(f_{p_{3/2}}/x)_{R=0}$ are
the $s$, $p_{1/2}$, and $(p_{3/2}/x)$ radial wavefunctions at zero,
and $m$ is the electron mass.
The terms included into the radial wavefunctions above are such
that the radial transition densities $U_{sp}=f_{s}f_{p}+g_{s}g_{p}$
for $s$-$p_{1/2}$ and $s$-$p_{3/2}$ include all corrections of
order $Z{^2}{\alpha ^{2}}$ and the lowest correction of order
$Z{^4}{\alpha ^{4}}$,
\begin{eqnarray}
\label{trdens1/2}
U _{sp_{1/2}}&=&\frac{2}{3}f_{s}(0)g_{p_{1/2}}(0)mR_{N}x\Big\{
1-\frac{3}{5}Z^{2}\alpha ^{2}x^{2}
\Big(
1-\frac{3}{14}x^{2}+\frac{2}{135}x^{4}
\Big)
+\frac{81}{560}Z^{4}{\alpha ^{4}}x^{4}
+...
\Big\} \\
\label{trdens3/2}
U _{sp_{3/2}}&=&f_{s}(0)\Big( f_{p_{3/2}}/x \Big)_{R=0} x
\Big\{
1-\frac{9}{20}Z^{2}{\alpha ^{2}}x^{2}
\Big(
1-\frac{69}{315}x^{2}+\frac{1}{63}x^{4}
\Big)
+\frac{243}{2800}Z^{4}{\alpha ^{4}}x^{4}+
... \Big\} \ .
\end{eqnarray}
It is seen by direct substitution of the transition densities
(\ref{trdens1/2},\ref{trdens3/2}) into the LDM expressions
(\ref{ldmextp},\ref{ldmcore},\ref{ldmdef}) that it is sufficient to
include in $U_{sp}$ just the first correction.
These terms ($\propto~Z^{2}{\alpha ^{2}}/R_{N}^{2}$) correspond
to the coefficient $b_{3}$ in expression (\ref{radwav}).
They give corrections to the Schiff moments of $\approx 15-25\%$.
The remaining terms in (\ref{trdens1/2},\ref{trdens3/2}) correct the
Schiff moments by a few percent.


\center
\widetext
\input psfig
\psfull

\end{document}